\documentclass[a4paper]{revtex4}

\usepackage{amsmath,amssymb}
\usepackage{epsfig}

\newcommand{\be}{\begin{equation}}
\newcommand{\ee}{\end{equation}}
\newcommand{\bea}{\begin{eqnarray}}
\newcommand{\eea}{\end{eqnarray}}
\newcommand{\gammac}{\gamma_c}
\newcommand{\f}{\frac}
\newcommand{\cN}{{\cal N}}

\begin{document}
\title{QCD traveling waves beyond leading logarithms}

\author{R. Peschanski}
\email{pesch@spht.saclay.cea.fr}
\affiliation{Service de physique th{\'e}orique, CEA/Saclay,
  91191 Gif-sur-Yvette cedex, France\footnote{
URA 2306, unit\'e de recherche associ\'ee au CNRS.}}
\author{S.~Sapeta}
\email{sapeta@th.if.uj.edu.pl}
\affiliation{M. Smoluchowski Institute of Physics, Jagellonian University, 
Reymonta 4, 30-059 Krak\'ow, Poland}

\begin{abstract}
We derive the asymptotic traveling-wave solutions of the nonlinear  1-dimensional
Balitsky-Kovchegov QCD equation for rapidity evolution in 
momentum-space, with 1-loop running coupling constant and equipped with the 
Balitsky-Kovchegov-Kuraev-Lipatov kernel at next-to-leading logarithmic 
accuracy, conveniently regularized by different resummation schemes. Traveling 
waves allow to  define ``universality classes'' of asymptotic solutions, i.e. 
independent of initial conditions and of the nonlinear damping. A dependence on 
the 
 resummation scheme remains, which is analyzed in terms of geometric scaling 
properties.
\end{abstract}

\maketitle


\section{Introduction}
\label{sec:intro}

In the  large-$N_c$ and ``mean-field'' approximation of high energy (high 
density) QCD,
the density of gluons with transverse momenta $k$ in a target evolves with 
rapidity 
$Y$ according to the nonlinear Balitsky-Kovchegov 
(BK) equation. This equation is supposed to capture essential features of 
saturation effects. In the 1-dimensional  approximation, and within leading 
logarithmic (LL) accuracy, it reads \cite{Balitsky:1995ub, Kovchegov:1999yj, 
Kovchegov:1999ua}
\be
\partial_Y \cN (L, Y) = 
\bar\alpha\, \chi (-\partial_L)\: \cN (L, Y) - \bar\alpha\, \cN^{\,2} (L, Y),
\label{1}\ee
where $L = \log (k^2/k^2_0)$, with $k^2_0$ being an arbitrary constant. 
In the LL approximation the characteristic function of the kernel has 
the standard  Balitsky-Kovchegov-Kuraev-Lipatov (BFKL) form
\cite{Lipatov:1976zz, Kuraev:1977fs, Balitsky:1978ic}, namely 
\be
\label{eq:bfkl_ll}
\chi(\gamma) = 2\psi(1) - \psi(\gamma) - \psi(1-\gamma),
\ee
and the coupling $\bar\alpha = \alpha_s N_c/\pi$ is kept
fixed. 

The goal of the present paper is to extend the known traveling-wave method 
\cite{Munier:2003vc, Munier:2003sj, Munier:2004xu} for obtaining asymptotic 
solutions of  the nonlinear equation \eqref{1} to the case where one 
considers the equation at the next-to-leading logarithmic (NLL) accuracy. In that 
case, 
the QCD 
coupling constant is running, i.e.
\be
\label{eq:b_def}
\bar\alpha(L) = \f{1}{bL}\ ,
\qquad \qquad
b = \f{11 N_c - 2 N_f}{12 N_c}\ ,
\ee 
and the 1-dimensional BK equation which we consider  reads 
\be
\label{eq:bk_nll}
bL\, \partial_Y \cN (L, Y) = 
\chi (-\partial_L, \partial_Y, \bar\alpha)\, \cN (L, Y) - \cN^{\, 2} (L, 
Y)\ .
\ee
As we shall recall later on, 
$\chi (-\partial_L, \partial_Y, \bar\alpha)$ is a ``renormalization-group 
improved ''  NLL 
kernel. It follows from specific resummation scheme implying higher order 
contributions while keeping the known expression 
\cite{Fadin:1998py,Ciafaloni:1998gs} of the NLL term in the 
kernel. Note that there could be  also NLL corrections to the nonlinear term, but, 
as we shall see from the traveling-wave properties, they do not change the 
asymptotic solutions.

To summarize the present situation of the QCD traveling waves for our purpose, it 
has been shown 
\cite{Munier:2003vc, Munier:2003sj, Munier:2004xu} that the BK equation with fixed 
coupling $\bar\alpha$ and in the LL approximation
belongs to the same universality class as the Fisher and 
Kolmogorov-Petrovsky-Piscounov (F-KPP) equation \cite{Fisher:1937, 
Kolmogorov:1937}. This means in particular that  the BK equation admits 
solutions in the form of {\it traveling waves} $\cN (L -\upsilon_g \bar\alpha 
Y)$. $L$ has the interpretation of a space variable while 
$t = \bar \alpha Y$, interpreted as time, is an increasing function of 
rapidity $Y.$  $\upsilon_g$ is the {\it critical velocity} of the wave, 
defined in this case as the minimum of the phase velocity. 
From the point of view of QCD the traveling-wave solution for the quantity 
$\cN$ translates into the property of geometric scaling \cite{geom} 
which yields 
$\cN (k^2/Q_s^2(Y))$, where the function $Q_s^2(Y) \propto \exp{(\upsilon_g 
\bar\alpha Y)}$ is called the saturation 
scale. Hence the asymptotic traveling-wave solutions of the BK equation satisfy 
geometric scaling (up to sub-dominant scaling violations which we will not analyze 
in the present work).

The existence of this solution depends on the general form of the  initial 
condition $\cN(L, Y_0)$  for large values 
of~$L$. 
It is however {\it universal}, i.e. independent on the 
details of the kernel or the specific form of 
the nonlinear term (e.g. independent of  NLL nonlinear terms) provided that it 
plays 
the r\^ole of a damping saturation term. This unique condition is that $\cN (L, 
Y_0)$ decreases at large $L$ at least as rapidly as an exponential $
\exp(-\gamma_0 Y)$ with $\gamma_0 > 
\gamma_c$, where $\gamma_c$ is the { critical} anomalous dimension, 
solution of the equation \be
\label{critical}
\chi (\gamma_c) = \gamma_c \, \chi' (\gamma_c)\ .
\ee 
The corresponding value for $\gamma_c $ obtained with the LL kernel 
(\ref{eq:bfkl_ll}) is $\gamma_c = 0.6275$ while the  value for the critical 
velocity $\upsilon_g = \chi' (\gamma_c) = 4.883$. 
The above condition is fulfilled for high energy QCD due to the ``color 
transparency'' property of the  gluon density, i.e.
$\cN (L, Y\!=\!Y_0) \sim e^{-L},$ valid at large $L$. Hence, $\gamma_0 = 1 
>\gamma_c\ 
.$

Within the framework of the traveling-wave approach the results for the gluon 
density $\cN (L, Y)$ and the saturation scale $Q^2_s (Y)$ were 
obtained in the LL approximation \cite{Munier:2003sj, Munier:2004xu}. 
Moreover, in \cite{Munier:2003sj}, the method was also 
extended to the case of the  
equation with the LL kernel and  1-loop QCD running coupling constant 
\eqref{eq:b_def}. In fact one considered the equation similar to 
\eqref{eq:bk_nll}, but keeping the LL kernel \eqref{eq:bfkl_ll}. In this case, 
the traveling-wave method leads to a different universality class of solutions 
than the F-KPP one, with, in particular, geometric scaling property in 
$\sqrt{Y}$ instead of~$Y.$

One would like to extend the above procedure the for BFKL kernel considered at the NLL 
accuracy and find the analytic solution in the framework of the traveling-wave 
approach for the corresponding BK 
equation \eqref{eq:bk_nll}. This has been done for the case of the fixed coupling $\bar\alpha$ \cite{enberg}. The case of the running coupling, which has not been considered so far, is the subject of this paper. 

As we know now, although the NLL corrections were calculated for the BFKL 
kernel 
\cite{Fadin:1998py, Ciafaloni:1998gs}
they turned out to be negative and so large that they are of no use for 
$\alpha_s$ if it is not extremely small.
However, several equivalent ways to cure this pathological behavior were 
proposed. They 
are based on the observation that the problem of the NLL corrections comes from 
the existence of spurious  collinear singularities.
This singularities may be canceled by resummation of the collinear terms at 
all orders satisfying, at the same time, the renormalization group 
constraints.

In practice, various resummation schemes were developed from which we will 
use the S3 and S4 schemes \cite{Salam:1998tj} and CCS scheme 
\cite{Ciafaloni:1999yw}. 
It is important for further use (and possible generalization to other 
schemes), to distinguish two types of resummation among the kernels. In the 
CCS ``implicit'' scheme \cite{Ciafaloni:1999yw}, the higher order 
resummation appears through the  dependence of the kernel on two 
variables  only $\chi (\gamma \!=\! -\partial_L, \omega \!=\!\partial_Y),$ 
where  $\omega \!= \!{\cal O}(\bar\alpha)$ drives the higher-order 
corrections. In the S3 and S4 
schemes \cite{Salam:1998tj}, an ``explicit'' dependence on the coupling 
constant 
appears, leading to a triple-variable dependence $\chi (\gamma \!=\! 
-\partial_L, \omega \!=\!\partial_Y, \bar\alpha)$. As we shall see later 
on, this introduces significant 
analytical (and eventually phenomenological) differences in the traveling 
wave solutions.

The aim of this paper is to apply the traveling-waves method with running 
coupling at the NLL level. As an outcome we obtain the result for the gluon 
density and the saturation scale valid at large $Y$ which is universal and can be 
used with any resummed NLL kernel. The specific cases of the S3, S4 and CCS 
kernels are studied in more detail.
We also compare the traveling-wave 
solutions with that obtained with a method of the linear BFKL evolution in the 
presence of absorbing boundary conditions \cite{Mueller:2002zm}, which has been 
applied using the CCS scheme \cite{Triantafyllopoulos:2002nz}.

The plan of the paper is the following. In Sec.~\ref{sec:bk_nll} we present in 
detail the calculation which leads to the solution of the BK equation with the 
NLL BFKL kernel in the case of running coupling. We arrive at the analytic 
asymptotic expressions for the saturation scale and the gluon density. This allows 
us to define in which universality class the traveling-wave solutions lie, 
depending on the type of the resummation scheme. In Sec.~\ref{sec:lambdas} we 
specifically analyze the dependence of the logarithmic derivatives of the 
saturation scale on the resummation scheme used.  We check the consistency 
and generality of our approach by comparison with  the previously known results 
for the CCS scheme.  
Finally, the conclusions and  outlook are given in Sec. \ref{sec:conclusions}.
\section{BK equation with NLL BFKL kernel and running coupling}
\label{sec:bk_nll}

Following the general method \cite{Munier:2003sj} we first write the solution 
to the linearized version of the BK 
equation at NLL~(\ref{eq:bk_nll}). It has the form  of double Mellin 
transform~\cite{Ciafaloni:1999yw}

\be
\label{eq:N}
\cN(L,Y) = \int \frac{d\gamma}{2\pi i}
         \int \frac{d\omega}{2\pi i} \,
         \cN_0(\gamma,\omega) \,
         \exp\left(-\gamma L + \omega Y + \frac{1}{b\omega} X(\gamma,\omega)
            \right),
\ee
with
\be
\label{eq:capital_x}
X(\gamma,\omega) = \int^{\gamma}_{\hat{\gamma}} 
                   d\gamma' \, \chi (\gamma', \omega)\ ,
\ee
and $\hat\gamma$ being an unspecified constant. Indeed, using the saddle-point 
method for the integration over $\gamma$ at large enough $L,$ one gets the  
saddle-point equation
\be
\label{eq:saddle}
-L+\frac{1}{b\omega}\ \chi (\gamma, \omega)=0\ ,
\ee
or equivalently in operator form
\be
\label{eq:consistency}
bL\, \partial_Y \cN (L, Y) = 
\chi (-\partial_L, \partial_Y)\, \cN (L, Y)\ ,
\ee
which is nothing else than the restriction of the BK equation \eqref{eq:bk_nll} to 
its 
linear part.

Here again $\chi$ as well as $X$ may depend on  $\bar\alpha$. In order to 
solve Eqs. (\ref{eq:N}) and (\ref{eq:capital_x}), we shall make the approximation of 
fixing $\bar\alpha$  in the kernel  at some, phenomenologically motivated, value, 
knowing that it applies only to the higher-order corrections. We shall come back 
to this approximation later on. In order  to simplify  the notation  we  do not 
write the  dependence on $\bar\alpha$ explicitly. 
Let us remind that the CCS kernel, 
as well as all other ``implicit'' schemes do not include this dependence on the 
coupling constant. 
 
As a next step we perform 
the saddle point integration over $\omega$ in the limit of large $Y$ and  
obtain
\be
\cN(L,Y) = \int \frac{d\gamma}{2\pi i}\ \cN_0(\gamma)
          \ \exp\Big(-\gamma L+F(\omega_s)\,Y\Big),
\ee
where 
$F(\omega_s)  = 
\frac{1}{Yb\ \omega_s}
\left(2 X(\gamma,\omega_s) - \omega_s \dot{X}(\gamma,\omega_s)\right)$ and the 
condition for the saddle point $\omega_s$ is given by the implicit 
equation
\be
\label{eq:cond2}
Yb\omega^2_s - X(\gamma,\omega_s) + \omega_s \dot{X}(\gamma,\omega_s) = 0\ .
\ee
We introduce the notation in which the ``dot'' means the derivative with 
respect to 
$\omega$ whereas the ``prime'' means the derivative with respect to $\gamma$.
Let us now expand the integral of the kernel \eqref {eq:capital_x}
 near $\omega =0$
\be
\label{eq:kernel_expansion}
X\,(\gamma, \omega)  =
 \sum^{\infty}_{p=0} \frac{X^{(p)}(\gamma,0)}{p!}\ \omega^{p}\ .
 \ee
Using a similar expansion for its  derivative $\dot{X}$ with 
respect to $\omega$ and substituting both quantities into Eq.~(\ref{eq:cond2}) we 
have 
\be
\label{sum}
\left[Yb + \frac{1}{2} \ddot{X}\, (\gamma, 0) \right] \omega^2_s =
X\, (\gamma, 0) - 
\left\{\sum^{\infty}_{p = 3} \frac{1}{p(p-2)!} X^{(p)}\, (\gamma, 0)\, \omega^p_s 
\right\}\ ,
\ee
by collecting  terms by powers of $\omega_s.$ It is clear from Eq.~\eqref{sum} 
that for asymptotic $Y,$ $\omega_s \sim Y^{-1/2}$ while in the convergence 
domain of the series the remaining terms between braces are at most of order 
$Y^{-3/2}.$  As we shall check later on, they may contribute only to 
non-universal sub-asymptotic terms. We also checked this by   looking for an 
iterative 
solution where the kernel
$\chi(\gamma, \omega)$ is expanded around $\omega_0$ and truncated at the 
order $P$. The universal terms are shown  not to depend on either  $\omega_0$ 
or the truncation.

\subsection{Traveling-wave critical parameters}

Considering then \eqref{sum} up to the order two, we obtain
\be
\label{l}\omega_s = 
\sqrt{
\frac{ X\,(\gamma,0) }{Yb + \frac{1}{2}\ddot{X}(\gamma,0)}}\ .
\ee
The saddle point value  behaves like 
$\omega_s \sim Y^{-\frac{1}{2}}$. With this form of $\omega_s$ the gluon 
density is given by
\be
\label{eq:ampl_int}
\cN(L,Y)  = \int \frac{d\gamma}{2\pi i} \cN_0(\gamma)  
\exp(-\gamma L + \Omega(\gamma) t),
\ee
where time is interpreted as
\be
\label{eq:time_def1}
t = \sqrt{Y + Y_0},
\ee
with $Y_0 = \ddot{X}(\gamma,0)/2b$. 
Interestingly,   this non-universal term in  formula \eqref{l}  absorbs the  
arbitrary 
constant $\hat\gamma$ in $\ddot{X}(\gamma,0).$ 
The dispersion relation has the form
\be
\label{eq:disp_rel}
\Omega(\gamma) = 
\sqrt{\frac{4}{b} X(\gamma,0) }\ .
\ee
In analogy to the LL case, following \cite{Munier:2003sj}, a {\it critical} group 
velocity 
(defined as the minimum of the {\it phase velocity} in the wave language) is 
obtained as 
\be
\label{eq}
\upsilon_g = 
\Omega(\gamma_c)/\gamma_c = \Omega'(\gamma_c)\ .
\ee
However, $\gammac$ determined in such a way still depends on the arbitrary 
constant $\hat\gamma$.
Thus,  requiring $\upsilon_g$ to be independent on the choice of $\hat\gamma$  
means imposing 
$d\upsilon_g(\hat\gamma)/d\hat\gamma = 0 = d\upsilon_g(\gamma_c)/d\gamma_c$.  
This is because the dependence of the velocity on $\hat\gamma$ comes through 
$\gamma_c$ only. 
Applying this condition to Eq.~\eqref{eq:disp_rel} one   gets
\be
\label{criticalbis}
d\upsilon_g(\hat\gamma)/d\hat\gamma 
=d\left(\Omega(\gamma_c)/\gamma_c\right)/d\gamma_c\ = 0 
\quad \Rightarrow \quad 
\upsilon_g\  =\ 
\sqrt{\frac{2\chi(\gammac,0)}{b\, \gammac}}\ ,
\ee
eliminating all dependence on the arbitrariness in the definition of $X$ in 
Eq.~\eqref{eq:capital_x}. This is to be contrasted with the situation for the 
linear problem with NLL kernels.

\begin{figure}[t]
\rotatebox{270}{\epsfig{file=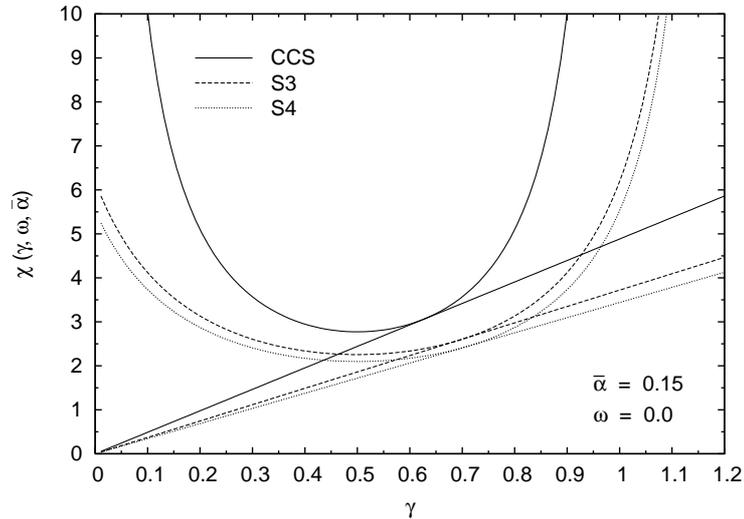,width=7.0cm}}
\caption{Graphical determination of the critical exponent $\gammac$ for the 
three resummed NLL kernels. The curve corresponding to the CCS scheme (at 
$\omega=0$) coincides with the LL curve.}
\label{fig:line}
\end{figure}

The value of $\gammac$ at the NLL level in general depends on the resummation 
scheme. This is clearly shown in  graphical form in 
Fig.~\ref{fig:line}. Geometrically, the  value of $\gammac$ is given by the 
tangent to the characteristic function of the kernel, different  for each NLL 
scheme. Note also that the curve corresponding to the CCS scheme at 
$\omega\!=\!0$ is nothing else than the LL curve given by  
Eq.~\eqref{eq:bfkl_ll}, and thus the critical parameters are the same in this 
case. This holds for any other ``implicit'' scheme which recovers the LL kernel at 
$\omega\!=\!0$ like e.g. those proposed in \cite{Kwiecinski:1996td, 
Andersson:1995ju} . 

\subsection{Asymptotic solution of the BK equation}

The linearized version of the BK equation (\ref{eq:bk_nll}) with the kernel 
expanded around $\omega_0$ up to the second order and the rapidity variable 
$Y$ changed to time variable $t=\sqrt{Y + Y_0}$ is given by 
\bea
\label{eq:bk_Lt}
\frac{bL}{2t}\, \partial_t \cN 
& = & \Big\{\overbrace{ -\frac{b}{2}\upsilon_g^2\, \partial_L + 
                        \frac{1}{2}\chi''(\, \partial_L^2 +
                        2 \gammac \, \partial_L +
                        \gammac^2)}^{\rm ``LL"}
\nonumber \\
&   & \underbrace{\ + \
      \frac{1}{2t} \, \dot{\chi}\, \partial_t  - \frac{1}{2t} \, \dot{\chi}'\, 
\partial_L \partial_t - \frac{1}{2t} \, \dot{\chi}'\, \gammac \partial_t +
      \frac{1}{8t^2} \,\ddot{\chi}\, (\partial_t^2 - \frac{1}{t}\, \partial_t)
      }_{\rm ``NLL"}
\Big\} \cN.
\eea
We singled out two parts in the above equation.
The part denoted as ``LL'' has already been present in the LL case (cf. 
Eq.~(33) from \cite{Munier:2003sj}). The remaining part called ``NLL'' contains 
new terms which originate from the dependence of the NLL BFKL kernel on 
$\omega$. 
In accordance with the approach developed in \cite{Munier:2003sj} we take the 
ansatz for the solution in the form \cite{Brunet:1997}
\be
\label{eq:ampl1}
\cN(z,t) = t^{\alpha}\, G\left(z\right)
         \exp\left(- \gammac\, z t^{\alpha}\right),
\qquad \qquad
z = \frac{L-\upsilon_g t+c(t)}{t^{\alpha}},
\ee
where we assume $\dot{c}(t) = \beta  t^{\,k-1}$ and the constants $\alpha$, 
$\beta$ and $k$ are to be determined. 
In order to recover the LL equation in the limit $\dot{\chi}, 
\dot{\chi}', \ddot{\chi} \to  0,$ we  set $\alpha = \frac{1}{3}$ and $k = 
\frac{1}{3}$, as in \cite{Munier:2003sj}. In fact this ansatz remains valid 
beyond the leading order. It is easy to check the consistency of this choice by 
looking at the time dependence of the new terms generated at higher order in 
Eq.~(\ref{eq:bk_Lt}). 
When all derivatives which appear in ``NLL'' part are calculated it turns out 
that their leading contributions are proportional to $t^{\frac{1}{3}}$. This 
means however that the terms in ``NLL'' part contribute only at the order 
$t^{-\frac{2}{3}}$ since each derivative is multiplied by at least the factor 
$t^{-1}$.
Thus, the BK NLL linearized equation in this approach has exactly the same 
form as in the LL case and reduces to the Airy equation
\be
\label{eq:airy2}
G\,''(z) = \frac{b \gammac \upsilon_g}{\chi''(\gammac, 0)} (z-4\beta)\, G (z).
\ee
The condition $G(z) \sim z$ as $z \to 0$ allows to fix the constant $\beta$ to 
\be
\beta = -\frac{1}{4} \left(\frac{\chi''(\gammac, 0)}{\gammac \, \upsilon_g\, 
b}\right)^{\frac{1}{3}}  \xi_1,
\ee
where $\xi_1 = -2.338$ is the zero of the Airy function.
Finally the result for the gluon density is given by 
\be
{\cal N}(L,t) = 
{\rm const} \cdot t^{\frac{1}{3}} \cdot {\rm Ai} \left(
\left(\frac{\sqrt{2 \gammac b \, 
\chi(\gammac,0)}}{\chi''(\gammac,0)}\right)^{\frac{1}{3}}
\ln\frac{k^2}{Q^2_s(t)} \; t^{-\frac{1}{3}} +
\xi_1 \right) \cdot
\left(\frac{k^2}{Q^2_s(t)}\right)^{-\gammac},
\ee
and the saturation scale up to a multiplicative constant has the form
\be
Q^2_s(t) = 
Q_0^2\, \exp \left(
\sqrt{\frac{2 \chi(\gammac, 0)}{b \gammac}}\; t + 
\frac{3}{4} 
\left(\frac{\chi''(\gammac,0)}{\sqrt{2 \gammac b \, 
\chi(\gammac,0)}}\right)^{\frac{1}{3}}\xi_1\; t^{\frac{1}{3}}\right).
\ee
Here $\chi(\gammac, 0)$ is the NLL BFKL kernel resummed in a given scheme (and 
possibly  taken at some value of $\bar\alpha$, see 
Sec.~\ref{sec:intro}). 
Hence, the solution of the BK equation with the resummed NLL kernel and running 
coupling has the same functional form as the solution for the LL kernel given 
in \cite{Munier:2003sj}. In particular, in the saturation scale the leading 
exponential term proportional to the time variable $t$ is supplemented by a 
universal  term 
in $t^{\frac{1}{3}},$ sub-leading by order $t^{-\frac{2}{3}}$. Note that the 
higher 
order 
contributions to the saddle-point Eq.~\eqref{eq:kernel_expansion}, being 
sub-leading 
by the
order $Y^{-\f 12}\sim t^{-1}$, do not interfere with this analysis and are 
expected 
to be non-universal.

Despite this formal similarity with the LL case with running coupling, the 
critical 
exponents are not the same for 
LL and NLL case if $\chi(\gamma, \omega\!=\!0, \bar\alpha)\ne \chi(\gamma),$ 
given in \eqref{eq:bfkl_ll}. This is due to the fact that the NLL result 
depends on the 
resummation scheme used for the kernel $\chi(\gamma, \omega\!=\!0, 
\bar\alpha)$. However, the traveling-wave solutions are not sensitive to the 
values of the kernels for $\omega\ne 0,$ contrary to the applications of the NLL 
BFKL linear 
equation without saturation (see, e.g. \cite{Peschanski:2004vw, Kepka:2006cg}).

\section{Saturation scales beyond leading order}

\label{sec:lambdas}

As we have seen, the time variable \eqref{eq:time_def1} scales like $t \sim 
Y^{1/2}.$ 
For a quantitative analysis of the solution, one defines the logarithmic 
time-derivative of 
the saturation scale  (or intercept) as
\be
\label{eq:lambdast_def}
\lambda_s^t (Y) = \frac{d \log (Q^2_s(Y)/\Lambda^2_{\rm QCD})}{dt}\ .
\ee
In order to compare with the usual definition of geometric scaling implying a 
linear rapidity 
dependence  of the saturation scale, one may also consider an ``effective'' 
intercept
\be
\label{eq:lambdas_def}
\lambda_s^{eff} (Y) = \frac{d \log (Q^2_s(Y)/\Lambda^2_{\rm QCD})}{dY}\ ,
\ee
so the relation between them is $\lambda_s^t(Y) = 2t \, \lambda_s^{eff}(Y)$.

\begin{figure}[t]
\rotatebox{270}{\epsfig{file=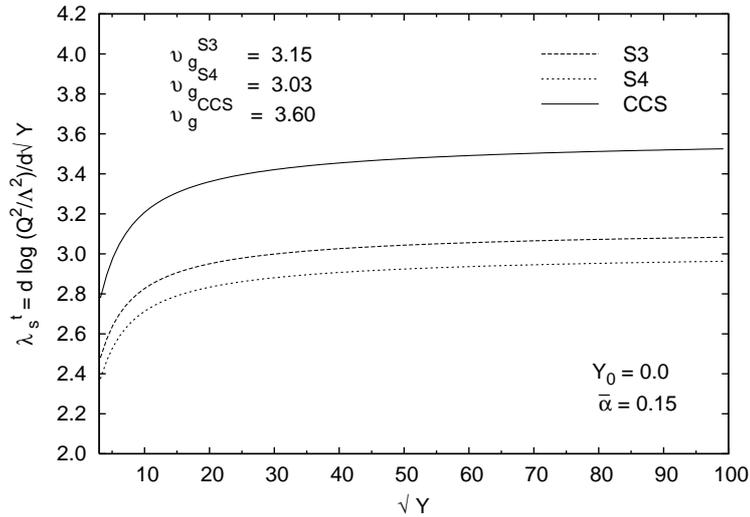,width=7.0cm}}
\caption{The logarithmic derivative $\lambda_s^t$ for various resummation 
schemes. We take $Y_0 = 0$ and $\bar\alpha = 0.15$.}
\label{fig:lst}
\end{figure}

From Eq.~(\ref{eq:lambdast_def}) we obtain 
\be
\label{eq:lambdast}
\lambda_s^t (Y) = 
\sqrt{\frac{2\chi(\gammac, 0)}{b \gammac}} +
\frac{1}{4} 
\left(\frac{\chi''(\gammac, 0)}
           {\sqrt{2\gammac b \chi(\gammac, 0)}}\right)^{\frac{1}{3}}\, 
\xi_1 \, t^{-\frac{2}{3}},
\ee
where the time $t$ is defined by Eq.~(\ref{eq:time_def1}). 
In Fig.~\ref{fig:lst} we show $\lambda_s^t$ from Eq.~(\ref{eq:lambdast}) as a 
function of ${Y}^{1/2}$ for the three different resummation schemes S3, S4 and 
CCS 
and the value of the coupling 
$\bar\alpha = 0.15$. The result depends on the scheme used. For $Y^{1/2} \to 
\infty$ the logarithmic derivative $\lambda_s^t$ goes to its asymptotic value 
equal to the group velocity $\upsilon_g$, which is also scheme-dependent.

\subsection{CCS scheme}

In fact, there appears a 
difference between the scheme CCS and the schemes S3 and S4. In the first 
case, since the higher order effects in the kernel are all given as a function 
of $\omega,$ one has identically $\chi(\gamma,\omega =0)\equiv \chi(\omega),$ 
i.e. the LL kernel of \eqref{eq:bfkl_ll}. In mathematical terms, this means 
that the CCS scheme falls into the same universality class of solutions as the 
equation with a LL kernel and running coupling constant. 

\begin{figure}[h]
\rotatebox{270}{\epsfig{file=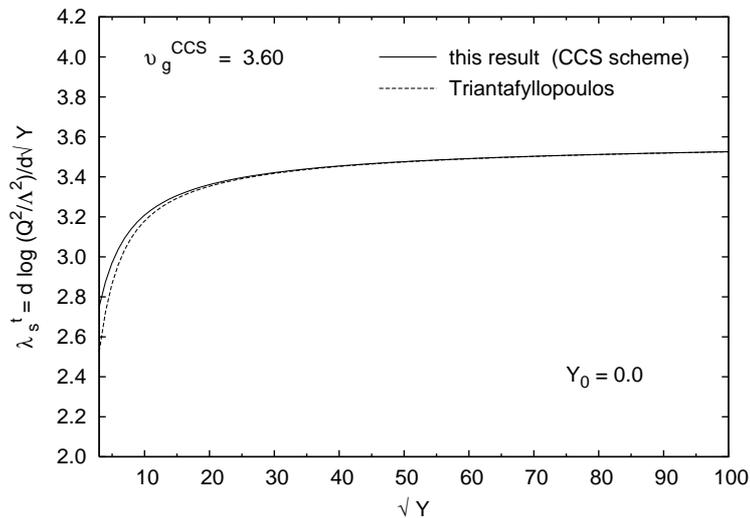,width=7.0cm}}
\caption{Comparison of the logarithmic derivative $\lambda_s^t$ obtained in 
the traveling-waves approach with the result of   
\cite{Triantafyllopoulos:2002nz}. For the sake of compatibility we use CCS 
scheme and $Y_0 = 0$. The corresponding formulas are given in 
Eqs.~(\ref{eq:lambdast}) and  (\ref{eq:lambdast_trian}).}
\label{fig:lstrian}
\end{figure}

In order to check this result, we compared the traveling-waves solution with 
that obtained with the different  method  \cite{Mueller:2002zm} applied to the 
CCS scheme \cite{Triantafyllopoulos:2002nz}. In this method, one considers 
only the linear evolution term 
supplemented by the absorbing  boundary conditions \cite{Mueller:2002zm}. It has  
been confirmed in \cite{Munier:2003vc, Munier:2003sj, Munier:2004xu} that the 
direct traveling-waves approach applied with the LL kernel leads to the same 
asymptotics (the traveling-wave method allows to add the third universal term 
\cite{Munier:2004xu} for the fixed coupling case while an eventual third term for 
the running case remains an open problem).

Extracting the asymptotic analytic form of  the solution found in  
\cite{Triantafyllopoulos:2002nz} (cf. Eqs. (58) and (59) therein) , and after 
changing it to our notations, one 
obtains
\be
\label{eq:lambdast_trian}
\lambda_s^t (Y) = 
\sqrt{\frac{2\chi(\gammac, 0)}{b \gammac}} +
\frac{1}{4} 
\left(\frac{\chi''(\gammac, 0)}
           {\sqrt{2\gammac b \chi(\gammac, 0)}}\right)^{\frac{1}{3}}\, 
\xi_1 \, 
\left(t - \sqrt{\frac{2b}{\gammac \chi(\gammac, 0)}}\right)^{-\frac{2}{3}}.
\ee
We see that the two results (\ref{eq:lambdast}) and (\ref{eq:lambdast_trian}) 
are consistent up to the corrections of the order $t^{-\frac{5}{3}}$. However, the 
corrections at this order are not expected to be  universal.
The detailed comparison of the two results for $\lambda_s^t$  given by 
Eqs.~(\ref{eq:lambdast}) and (\ref{eq:lambdast_trian}) is shown in 
Fig.~\ref{fig:lstrian}. 
We observe that they converge for the asymptotic values of $Y$.

\begin{figure}[hb]
\rotatebox{270}{\epsfig{file=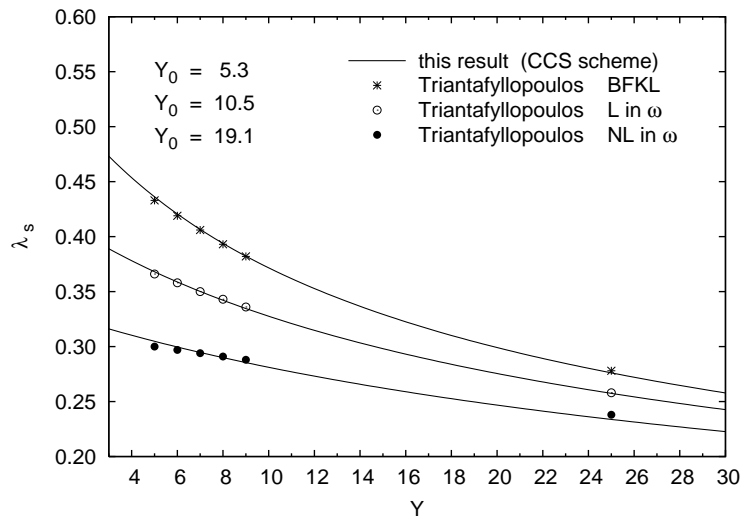,width=7.0cm}}
\caption{Comparison of the logarithmic derivative $\lambda_s$ obtained in the 
traveling-waves approach in the CCS scheme and various $Y_0$ values with the 
results of \cite{Triantafyllopoulos:2002nz}. The notations BFKL ``L in 
$\omega$'', ``NL in $\omega$'' correspond to the running coupling with the LL 
kernel and two different treatments of the CCS scheme, respectively, see 
\cite{Triantafyllopoulos:2002nz} for details.}
\label{fig:ls}
\end{figure}

For a  quantitative comparison at lower rapidities, we can also confront our 
result for $\lambda_s^{eff}$ with that given in 
\cite{Triantafyllopoulos:2002nz}. This is presented in Fig.~\ref{fig:ls}. We 
see that $\lambda_s^{eff}$ based on the definition of 
Eq.~(\ref{eq:lambdas_def}) 
with appropriate $Y_0$ can successfully mimic the 
results from~\cite{Triantafyllopoulos:2002nz}. In our analysis, varying $Y_0$
plays the r\^ole of parameterizing typical non-universal terms, i.e. terms 
which depend on the initial conditions, details of the kernel, or of the method 
used for extracting the asymptotic behavior. This  is a hint for  the 
compatibility of the results of \cite{Triantafyllopoulos:2002nz} with the 
universality class defined by the LL kernel with running coupling constant. The 
analysis of non-universal terms is beyond the scope of the present paper and 
deserves 
further study, since these terms may be phenomenologically important.

\subsection {S3 and S4 schemes}

\begin{figure}[ht]
\rotatebox{270}{\epsfig{file=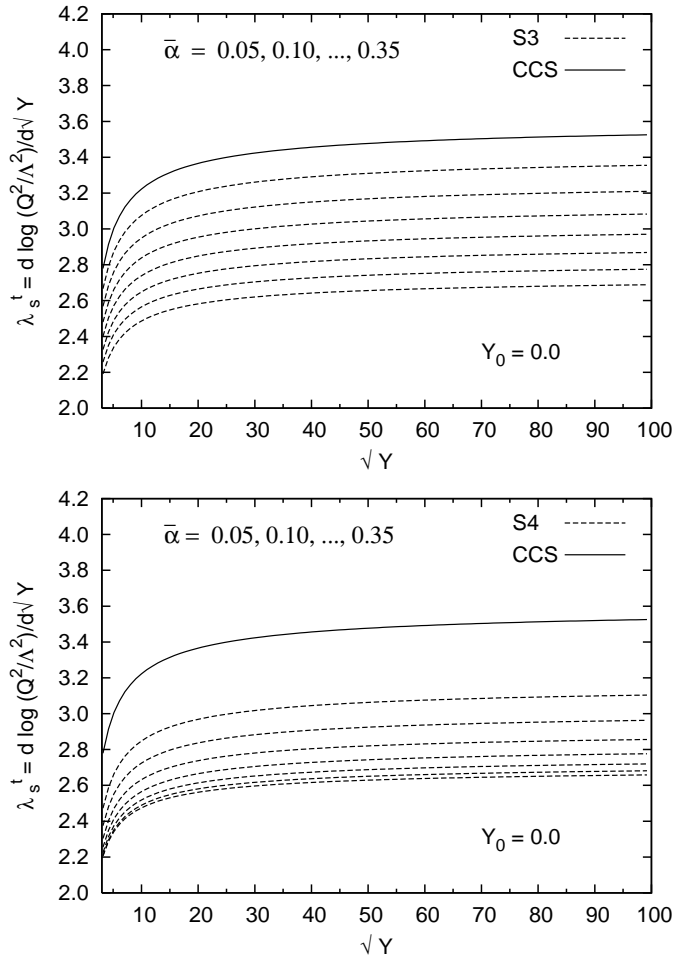,width=9.0cm,angle=90}}
\caption{Dependence of the logarithmic derivative $\lambda_s^t$ obtained in the 
traveling-waves approach in the S3 and S4  schemes as a function of 
$\bar\alpha.$ Top: S3 scheme; Bottom: S4 scheme. The curves go up when the value of $\bar\alpha$ decreases. The fixed curve corresponding 
to the CCS scheme, i.e. the same as for LL kernel, is also indicated in both 
cases.}
\label{fig:lsu}
\end{figure}

The schemes S3 and S4, by contrast, are not expected to lie in the same 
universality class as the previous one. Indeed, the resummed kernels depend 
explicitly on the  the value of the coupling constant.  This is depicted in 
Fig.~\ref{fig:lsu}, where one can see how
the time-derivative of the saturation scale varies 
with $\bar\alpha.$ 
In order to characterize the universality class of these schemes 
one would have   to go one step beyond the approximation of a fixed  $\bar\alpha$ 
{\it in the kernel} that we used in the beginning. In fact one would have to 
first solve exactly the linear operator equation
\be
\label{34l}
bL\, \partial_Y \cN (L, Y) = 
\chi (-\partial_L, \partial_Y, \bar\alpha(L)\!=\!1/{bL})\ \cN (L, Y) \ ,
\ee
and then apply the general traveling-waves method to find the critical velocity and 
asymptotic solution. We postpone this detailed study for future work. What, however,  
is indicated by our study is that the ``universality class'' of the solution may 
be different from the CCS case. It remains to be found which kind of  
geometric scaling property will be satisfied.

Hence we obtain the new result from the QCD traveling-waves approach that 
the specific  asymptotic solutions of 
the BK equation at NLL accuracy depend on the resummation schemes. In particular 
they vary between the schemes which include an  {\it explicit} or {\it implicit} 
dependence on the running coupling constant in the definition of the NLL kernel. 
This 
introduces a theoretical distinction between NLL effects. It would be an 
interesting issue to know whether this distinction may appear at physically 
reachable rapidities. This deserves certainly more study in the future.

\section{Conclusions}
\label{sec:conclusions}

We considered the 1-dimensional Balitsky-Kovchegov (BK) equation in momentum 
space with running coupling constant. The  Balitsky-Kovchegov-Kuraev-Lipatov 
(BFKL) kernel was taken at the next-to-leading logarithmic (NLL) accuracy with 
higher orders following known resummation schemes and satisfying 
renormalization-group (RG) constraints. 

Using mathematical properties of nonlinear equations, we 
derived the traveling-wave solutions, which are valid in the limit of large 
rapidities Y and obey universality properties, i.e. are independent on the 
specific form of the initial conditions, the detailed form of the kernel and  the 
nonlinearities.
 
We found that the results for the gluon density and the saturation scale 
acquire dependence on the resummation scheme. For a scheme with an {\it 
implicit} higher order dependence, i.e. where  the NLL kernel
 is a function of $\gamma\!=\!-\partial_L$ and $\omega\!=\! \partial_Y$ only, 
 one is expected to fall into the same universality class  as 
in the LL, running coupling case. 
This is explicitly derived for the CCS scheme \cite{Ciafaloni:1999yw}.

For schemes with an {\it 
explicit} higher order dependence, such as S3 and S4 of 
\cite{Salam:1998tj}, where in addition the NLL kernel is an explicit function of 
$\bar\alpha$, the asymptotic solution does change as a function of the value of 
$\bar\alpha$. The precise 
determination of the universality class when $\bar\alpha$ is also considered as 
running in the NLL kernel is an interesting challenge for the future work.

It is interesting to address further questions about the properties of QCD 
saturation beyond leading logarithms. For instance, it will be useful to ask 
which is the universality class of the BK equation (or whether they are 
modified) when different ways of implementing the running coupling constant 
are used, since there remain an ambiguity concerning this problem (see e.g. 
\cite{Kovchegov:2006vj,Balitsky:2006wa}). On a more phenomenological 
ground, one would like to say something about the physically reachable  
rapidity region where non-universal terms may be important. For instance, a 
way was proposed in \cite{Peschanski:2005ic, Marquet:2005ic} to take into 
account the actual form of nonlinearities beyond the asymptotics where their 
specific form does not play a big r\^ole (except in the selection of the 
critical velocity by setting the unitarity limit). Finally, there remains the 
problem of a 
more complete QCD solution with NLL accuracy going beyond the mean-field 
approximation. We postpone these studies for further work.

\begin{acknowledgments}
R.P. thanks Cyrille Marquet and Gregory Soyez for useful suggestions. S.S. is 
grateful to S. Munier and K.~Golec-Biernat for valuable discussions and 
acknowledges grants of the Polish Ministry of Education and Science: 
No.~1~P03B~028~28 (2005-08) and N202~048~31/2647~(2006-08). 
\end{acknowledgments}


\end{document}